\documentclass[conference]{IEEEtran}
\IEEEoverridecommandlockouts
\usepackage{cite}
\usepackage{amsmath,amssymb,amsfonts}
\usepackage{algorithmic}
\usepackage{graphicx}
\usepackage{textcomp}
\usepackage{xcolor}
\usepackage{algorithmic}
\usepackage{algorithm}
\usepackage{hyperref}
\usepackage{soul}
\newcounter{mycounter}

\def\BibTeX{{\rm B\kern-.05em{\sc i\kern-.025em b}\kern-.08em
    T\kern-.1667em\lower.7ex\hbox{E}\kern-.125emX}}
   
\DeclareRobustCommand{\IEEEauthorrefmark}[1]{\smash{\textsuperscript{\footnotesize #1}}}    

\begin{document}
\bstctlcite{IEEEexample:BSTcontrol}

\title{Sum Rate Maximization in STAR-RIS Assisted Full-Duplex Communication Systems\\
}

\author{\IEEEauthorblockN{Pulasthi P. Perera\IEEEauthorrefmark{1}, Vanodhya G. Warnasooriya\IEEEauthorrefmark{1}, Dhanushka Kudathanthirige\IEEEauthorrefmark{2}, and Himal A. Suraweera\IEEEauthorrefmark{1}} 
	\IEEEauthorblockA{\IEEEauthorrefmark{1}Department of Electrical and Electronic Engineering, University of Peradeniya, Sri Lanka\\
	\IEEEauthorrefmark{2}
	Department of Physics and Engineering, Cornell College, Mount Vernon, IA, USA, 52314\\Email: \href{pulasthiperera@ieee.org,vanowarna@ieee.org}{\{pulasthiperera, vanowarna\}@ieee.org}, \href{dkudathanthirige@cornellcollege.edu}{dkudathanthirige@cornellcollege.edu},
	\href{himal@ee.pdn.ac.lk}{himal@ee.pdn.ac.lk}
	\vspace{-1mm}}
}

\maketitle

\begin{abstract}
The sum rate performance of simultaneous transmitting and reflecting reconfigurable intelligent surface (STAR-RIS) assisted full-duplex (FD) communication systems is investigated. The reflection and transmission coefficients of STAR-RIS elements are optimized for the energy splitting and mode switching protocols to maximize the weighted sum rate of the system. The underlying optimization problems are non-convex, and hence, the successive convex approximation technique has been employed to develop efficient algorithms to obtain sub-optimal solutions. Thereby, the maximum average weighted sum rate and corresponding coefficients at the STAR-RIS subject to predefined threshold rates and unit-modulus constraints are quantified. The performance of the proposed system design is compared with the conventional reflecting/transmitting-only RISs and half-duplex counterparts via simulations where it is observed that STAR-RIS can boost the performance of FD systems.  

\end{abstract}

\vspace{-2mm}
\section{Introduction}

Owing to the recent advancements of meta-materials/surfaces, reconfigurable intelligent surface (RIS) has emerged as a promising technology for the implementation of sixth generation wireless networks \cite{Basar2019,Wu2019}. By utilizing a large number of passive elements, a RIS intelligently controls the phase shifts of incident signals to satisfy the requirements of various wireless applications \cite{Basar2019,Wu2019,Chen2019,Kudathanthirige2021}. 

Most of the existing research considers RIS as a reflecting-only surface that can only guide the incident signals into the space in front of it  \cite{Chen2019,Kudathanthirige2021}. Therefore, the transmitter and receiver need to be placed on the same side of the RIS \cite{Basar2019,Wu2019,Chen2019,Kudathanthirige2021}, and any potential receiver on the other side of the RIS will not be able to communicate with the transmitter. However, in practice, it is most likely to have users on both sides of the RIS.  In order to overcome this topological restriction, a novel concept called simultaneously transmitting and reflecting RISs (STAR-RISs) was recently  proposed in \cite{Xu2021,Yuanwei2021,Wu2021}. In particular, the elements of a  STAR-RIS are capable of reflecting and transmitting  the incident signals into the spaces in front and behind the RIS, respectively, covering full space. Both transmitted and reflected signals can be controlled  by adjusting the transmission and reflection coefficients of each passive element via three operating protocols; namely, Energy Splitting (ES), Mode Switching (MS), and Time Switching (TS) \cite{Xu2021,Yuanwei2021}.  On a parallel development, full-duplex (FD) communication in which two or more users concurrently exchange information within the same time-frequency resource block has received extensive attention as it can double the spectrum utilization \cite{Riihonen2011,Kolodziej2019}. 

\noindent\textbf{Prior related research:} Several research contributions have investigated the integration of RIS technology with FD two-way communication  \cite{Atapattu2020,Zhang2021,Cai2021,Xu2021b,Peng2021}. References \cite{Atapattu2020} and  \cite{Zhang2021} investigate sum rate maximization of a RIS assisted two-way wireless network with single-input single-output users and multi-antenna users, respectively. In \cite{Cai2021}, transmit power minimization of RIS assisted FD systems is considered by jointly optimizing passive reflection coefficients at the RIS, transmit powers at access point (AP), and uplink users.
A cognitive system with multiple half-duplex (HD) users and an FD secondary  AP is considered in \cite{Xu2021b}, where the secondary sum rate is maximized subject to primary interference constraints. In \cite{Peng2021}, the  RIS phase shift coefficients are optimized to maximize the minimum user rate and mitigate the inter-user interference. Reference \cite{Sharma2021} explores the RIS technology for minimizing the effect of self-interference (SI), which can cause performance degradation in FD communication systems \cite{Riihonen2011}. 

\noindent\textbf{Motivation and our contribution:} 
In \cite{Atapattu2020,Zhang2021,Cai2021,Xu2021b,Peng2021}, a RIS is introduced to supplant the operation of an FD amplify-and-forward relay since a RIS can generate fine-grained 3-dimensional beamforming vectors as that of a multi-antenna relay. However, unlike a relay, a RIS passively modifies the existing signals without transmitting any new ones, eliminating any kind of signal processing requirements for SI cancellation at the RIS. Hence, RIS improves both energy efficiency and the spectral efficiency of FD communication system \cite{Cai2021}. 

\begin{figure}[!t]\centering \vspace{0mm}
 	\def\svgwidth{230pt} 
 	\fontsize{8}{4}\selectfont 
 	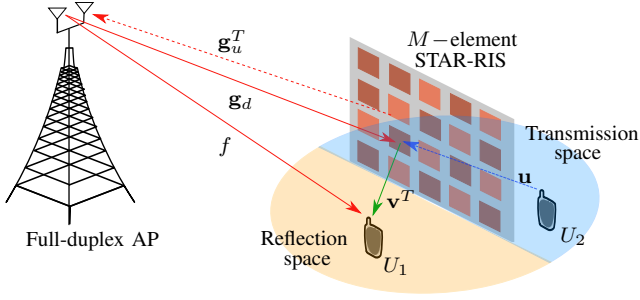 \vspace{0mm}
 	\caption{A STAR-RIS assisted FD wireless communication setup.}\vspace{-6mm} \label{fig:system_model}
 \end{figure}

Since STAR-RIS outperforms the conventional RIS in terms of flexibility and coverage \cite{Xu2021,Yuanwei2021,Wu2021}, in this paper, we investigate the viability of  STAR-RIS assisted FD communication systems. 
 As depicted in Fig. \ref{fig:system_model}, a setup in which an FD AP communicates with two HD users via a STAR-RIS is assumed.
Users are located on the opposite sides of the STAR-RIS and concurrently operate in uplink and downlink. In order to improve the performance, the reflection and transmission coefficients of the RIS are optimized for ES and  MS operating protocols to maximize the weighted sum rate. However, the resulting optimization problems are non-convex. Hence, low complexity sub-optimal solutions are proposed based on successive convex approximation (SCA) techniques. Our numerical results reveal that the STAR-RIS architecture can successfully be integrated into FD systems to boost the communication performance.


\noindent \textbf{Notation:} $\mathbf x^T$ and $\mathbf x^H$ denote the transpose and conjugate transpose of a vector $\mathbf x$. Moreover, $\mathrm{diag} (\mathbf x)$ generates a diagonal matrix with elements of the vector $\mathbf x$ as diagonal elements. The real part and phase of a complex number $z$ are given by $\text{Re}[z]$ and $\text{Arg}(z)$, respectively. $Z$ $\sim$  $\mathcal {CN}\left( 0,1 \right) $ denotes that $Z$ is the circularly symmetric Gaussian distributed random variable with zero mean and unit variance.

\section{System, Channel, and Signal Models}
\subsection{System and Channel Models}

We consider an FD communication system in which an FD AP exchanges information with two single-antenna HD user nodes (namely, $U_1$ and $U_2$) through an $M$-element STAR-RIS as shown in Fig. \ref{fig:system_model}. 
The AP has two antennas, one each for signal transmission and reception as in \cite{Cai2021}. 
The $U_1$ and $U_2$ are located in the reflection space and transmission space of the STAR-RIS, respectively. 
Without loss of generality, the AP transmits a signal to $U_1$ while receiving a signal from $U_2$. The configurations of the RIS can be intelligently controlled by a micro-controller, which gets the necessary instructions from the AP over a high-speed, error-free wired or wireless back-haul link \cite{Atapattu2020}.


The AP-to-RIS  and RIS-to-AP channels are denoted by $\mathbf g_d\in \mathbb C^{M\times 1}$ and  $\mathbf g^T_u\in \mathbb C^{1\times M}$, respectively. The channels from RIS-to-$U_1$ and $U_2$-to-RIS are denoted by $\mathbf v^T \in \mathbb C^{1\times M}$ and $\mathbf u\in \mathbb C^{M\times 1}$. 
 Moreover, the SI channel between the transmit and receive antenna is represented by $h_{AP}$. The $m$th element of these channel vectors can be modeled as $q_m =  \bar{q}_m e^{-j \psi_{q,m}}$, where $\bar{q}_m$ is the envelope and $\psi_{q,m}$ is the phase of channel coefficient $q_m$  for $q\in\{g_u, g_d,u,v\}$ and $m\in \mathcal M$, where $\mathcal M =\{1,\ldots, M\}$. Similarly, the AP-to-$U_1$ direct channel is denoted by $f = \bar{f }e^{-j\psi_f}$ and  the channel between $U_2$ and AP is assumed to be blocked. The envelope and phase of each channel coefficient are modeled as Nakagami-$m$ and uniform random variables, respectively. In particular, $\bar{q}_m\sim \mathrm{Nakagami}\,(m_q, m_q \zeta_q),\psi_{q,m}\sim \mathrm{Uniform}\left[\right.\!\!0, 2\pi\!\left.\right)$ for $m\in \mathcal M$ and $\bar{f} \sim \mathrm{Nakagami}\,(m_f, m_f \zeta_f),
\psi_{f}\sim \mathrm{Uniform}\left[\right.\!\!0, 2\pi\!\left.\right)$. Here, $\zeta_q$ and $\zeta_f$ account for the large-scale fading/path loss experienced by the channels $q_m$, $\forall m \in \mathcal M$ and $f$, respectively. It is assumed that the channel coefficients experience independent fading and are perfectly known at the AP.

The transmission and reflection properties of the $m$th RIS element are given by $ \sqrt{\beta_{m}^{t}} e^{j\theta^{t}_{m}}$ and $ \sqrt{\beta_{m}^{r}} e^{j\theta^{r}_{m}}$, where $\beta_m^t, \beta_m^r\in [0,1]$ and $\theta_m^t, \theta_m^r\in \left[\right.0,2\pi\left.\right)$ characterize the amplitude and phase modifications imposed on the transmitted and reflected signal components, respectively. The phase modifications can be generally configured independent of each other, however, the amplitude coefficients are constrained by the law of energy conservation such that \cite{Yuanwei2021} \begin{eqnarray}\label{eqn:beta_constraint}
	\beta_{m}^{t}+\beta_{m}^{r}=1, \quad \forall m \in \mathcal M.
\end{eqnarray}
By exploiting the inherent flexibility of the RIS elements, we consider ES and MS operating protocols at the STAR-RIS. According to the ES protocol, all elements concurrently transmit and reflect the incident signals. In the MS protocol, each of the STAR-RIS element operates either in the full transmission mode (i.e., $\beta^t_m = 1$, $\beta^r_m = 0$) or full reflection mode (i.e., $\beta^t_m = 0$, $\beta^r_m = 1$). 
The sets of $2M$ (both transmit and reflection) coefficients are  concisely represented as vectors $\mathbf s_{\text{x}}^{r} = \left[ \sqrt{\beta_1^r}e^{j\theta^r_1}, \cdots,
\sqrt{\beta_m^r}e^{j\theta^r_m},\cdots,\sqrt{\beta_M^r}e^{j\theta^r_M}\right]^T \in \mathbb C^{M\times 1}$, $\mathbf s_{\text{x}}^{t} = \left[ \sqrt{\beta_1^t}e^{j\theta^t_1}, \cdots,\sqrt{\beta_m^t}e^{j\theta^t_m},\cdots,\sqrt{\beta_M^t}e^{j\theta^t_M}\right] ^T$ $\in \mathbb C^{M\times 1}$, or equivalently in matrix form as $\mathbf \Theta^r_\text{x} = \mathrm{diag}\left(\mathbf s_{\text{x}}^{r}\right)$  and  $\mathbf \Theta^t_\text{x} = \mathrm{diag}\left(\mathbf s_{\text{x}}^{t}\right)$, where $\text{x} \in \{\text{E},\text{M}\}$ indicates the STAR-RIS operating protocol.


\subsection{Signal Model}
The AP concurrently transmits and receives signals to $U_1$ and from $U_2$, respectively. 
The received signal at $U_1$ can be written as 
\begin{eqnarray}\label{eqn:user_1_signal}
	y_{\text{x},1} = \sqrt{P_1} (f+\mathbf v^T \boldsymbol{\Theta}^r_{\text{x}} \mathbf g_d)x_1
	+ \sqrt{P_2}\mathbf v^T\boldsymbol{\Theta}^t_{\text{x}} \mathbf u x_2 +n_1,
\end{eqnarray}
where $P_i$ is the transmit power of the signal $x_i$, which satisfies $\mathbb E\{|x_i|^2\}=1$ for  $i\in\{1,2\}$. Here, $x_1$ and $x_2$ are signals transmitted from the AP and $U_2$, respectively.
The first term in (\ref{eqn:user_1_signal}) captures the desired signal received at $U_1$ through direct and reflected paths via the STAR-RIS. The second term accounts for the inter-user interference. 
Moreover, in (\ref{eqn:user_1_signal}), $n_1$ is the additive white Gaussian noise (AWGN) distributed as $n_1 \sim \mathcal {CN}(0, \sigma^2_1)$. The signal-to-interference-plus-noise ratio (SINR) of the $U_1$ can be written as 
\begin{eqnarray}\label{eqn:gamma1} 
	\gamma_{\text{x},1} &=& \frac{{\bar{\gamma}}_1\left|f+\mathbf v^{T}  \boldsymbol{\Theta}^r_{\text{x}} \mathbf g_d\right|^{2}}{{\bar{\gamma}'_2}\left|\mathbf v^{T}\boldsymbol{\Theta}^t_{\text{x}}\mathbf u\right|^{2}+1} ,
\end{eqnarray}
where $\bar{\gamma}_1 = P_1/\sigma_1^2$ and $\overline{\gamma}'_2= P_2 / \sigma_1^2$.

Next, the signal received by the AP can be written as
	\begin{eqnarray}\label{eqn:user_2_signal}
    y_{\text{x},2} \!=\! \sqrt{\!P_2} \mathbf g^T_u\boldsymbol{\Theta}^t_{\text{x}} \mathbf u x_2
    \!+\! \sqrt{\!P_1}
    \left( h_{AP} + \mathbf g^T_u \boldsymbol{\Theta}^r_{\text{x}}\mathbf g_{d}
    \right) x_1 \!+\! n_{2},
\end{eqnarray}
where the first term is the desired signal and $n_2 \sim \mathcal {CN}(0, \sigma^2_2)$ is the AWGN.  The second term characterizes the interference, which is a cumulative effect of SI occurring due to downlink transmission at the AP  and the reflection of the downlink signal via the STAR-RIS. The former can be partially mitigated using the existing  SI mitigation methods \cite{Riihonen2011}. The latter is negligible due to severe propagation attenuation when the RIS is placed near to the users \cite{Meng2021}.
Thus, any residual effect generated from SI cancellation is modeled as $ g_{\text{AP}} \sim \mathcal{CN}(0, \sigma^2_{SI})$, where $\sigma^2_{SI}$ is the average power of the SI. 
Then, (\ref{eqn:user_2_signal}) reduces to
\begin{eqnarray} \label{eqn:user_2_modified}
	\hat{y}_{\text{x},2} &=& \sqrt{P_2} \mathbf g^T_u\boldsymbol{\Theta}^t_{\text{x}} \mathbf u x_2+ \sqrt{P_1}g_{\text{AP}}x_1 + n_2.
\end{eqnarray} 
\noindent By using \eqref{eqn:user_2_modified}, SINR at the AP can be derived as
\begin{eqnarray}\label{eqn:gamma2}
   \gamma_{\text{x},2} = \bar{\gamma}_2\left|\mathbf g^{T}_u \boldsymbol{\Theta}^t_{\text{x}} \mathbf u \right|^{2},
\end{eqnarray}
where $\bar{\gamma}_2 = P_2/\sigma_2^2 (P_1|g_{\text{AP}}|^2/\sigma_2^2+1)^{-1}$. To facilitate STAR-RIS design, 
we let $ (\mathbf s_{\text{x}}^{r})^T \mathbf h = \mathbf v^T\mathbf {\Theta}_\text{x}^{r} \mathbf g_d$,
$	(\mathbf s^{t}_{\text{x}})^T\mathbf q=\mathbf v^{T}\mathbf\Theta_\text{x}^{t}\mathbf u$, and  
$(\mathbf s_{\text{x}}^{t})^T \mathbf z  =	\mathbf g_u^{T} \mathbf \Theta_\text{x}^{t} \mathbf u$, where 
$\mathbf h= \mathrm{diag}(\mathbf v^T) \mathbf g_d$, $ \mathbf q = \mathrm{diag} (\mathbf v^T) \mathbf u$, and  $ \mathbf z = \mathrm{diag} (\mathbf g_u^T) \mathbf u$. Using (\ref{eqn:gamma1}) and \eqref{eqn:gamma2}, corresponding rates of the users can be defined as
\begin{eqnarray}
	\mathcal R_{\text{x},1} \! &=& \! \!
	\log_2 \left(\! 1 + \frac{\bar{\gamma}_1\left|f+(\mathbf s^r_\text{x})^T  \mathbf h \right|^{2}}
	{{\bar{\gamma}'_2}\left|(\mathbf s^t_\text{x})^T\mathbf q\right|^{2}+1}\right)\label{eqn:user_1_rate},  \text{ and }\\
\mathcal R_{\text{x},2} &=& \log_2  \left(1+\bar{\gamma}_2\left|(\mathbf s^t_\text{x})^T \mathbf z \right|^2 \right).\label{eqn:user_2_rate}
\end{eqnarray}

\subsection{Problem Formulation}

 The transmission and reflection coefficients of the RIS can be optimized to ensure constructive signal combining at the AP and $U_1$. Therefore, we consider a weighted sum rate maximization problem as follows:
 \begin{subequations}
 	\begin{eqnarray}\label{eqn:OptimizationProblem}
 	 	\mathrm {P}1:\;	\underset{\mathbf s^k_{\text{x}}\;\; \forall k}
 		{\text{maximize}}&& \!\!\!\! w_1 \mathcal R_{\text{x},1} + w_2 \mathcal R_{\text{x},2}\label{eqn:ObjectiveP1}\\
 		\text{subject to} && \nonumber\\
 		&&\!\!\!\!\!\!\!\!\!\!\!\!\!\!\!\!\!\!\!\!\!\!\!\!\!\!\!\!\!\!\!\!\!\!\!
 		\mathrm C_{1}: 0\leq \text{Arg}\left(s^k_{\text{x},m}\right)< 2\pi, \quad  \forall m \in \mathcal M,
 		\label{eqn:phaseconstraint} \\
 		&&\!\!\!\!\!\!\!\!\!\!\!\!\!\!\!\!\!\!\!\!\!\!\!\!\!\!\!\!\!\!\!\!\!\!\!
		\mathrm  C_{2}:
		|s^r_{\text{x},m}|^2 + |s^t_{\text{x},m}|^2 = 1,\quad \forall m \in \mathcal M, \label{eqn:amplitudeconstraintxx}\\
 		&&\!\!\!\!\!\!\!\!\!\!\!\!\!\!\!\!\!\!\!\!\!\!\!\!\!\!\!\!\!\!\!\!\!\!\!
 		\mathrm  C_{3}:
		0\leq |s^k_{\text{x},m}| \leq 1, \;\;\forall m \in \mathcal M, \text{ for ES,} \label{eqn:amplitudeconstraint} \\
		&&\!\!\!\!\!\!\!\!\!\!\!\!\!\!\!\!\!\!\!\!\!\!\!\!\!\!\!\!\!\!\!\!\!\!\!
		\mathrm  C_{4}:
		 |s^k_{\text{x},m}| \in \{0,1\},\; \forall m \in \mathcal M \text{ for MS,}\label{eqn:amplitudeconstraint2} \\
		&&\!\!\!\!\!\!\!\!\!\!\!\!\!\!\!\!\!\!\!\!\!\!\!\!\!\!\!\!\!\!\!\!\!\!\!
		\mathrm  C_{5}:
		\mathcal R_{\text{x},1} \geq \mathcal R^{th}_1 \text{ and }  \mathcal R_{\text{x},2} \geq \mathcal R^{th}_2,
		\label{eqn:rateconstraints}	
 	\end{eqnarray}
 \end{subequations} 
where $\forall k \in \{r,t\}$. The constraints $\mathrm C_{1}$-$\mathrm C_{4}$  model the phase and amplitude limitations of the STAR-RIS elements.  $\mathrm C_{3}$  and  $\mathrm C_{4}$ are  amplitude constraints for the ES and MS protocols, respectively. In (\ref{eqn:rateconstraints}), $\mathcal R^{th}_1$ and $\mathcal R^{th}_2$ are the minimum quality-of-service (QoS) requirements  of $U_1$ and $U_2$, respectively. Moreover, $w_1$ and $w_2$ are the weighting factors such that  $0\leq w_1,w_2 \leq 1$ and $w_1+w_2 =1$. These factors  represent the relative priority of uplink/downlink user rates. To be specific, a large weight factor leads to a higher priority.  Our formulation is well suited for modeling the asymmetric uplink and downlink rates in a typical cellular system.

\noindent \textbf{Remark 1:} In   	$\mathrm {P}1$, the objective is to maximize weighted sum rate by controlling the transmit and receive coefficients at the STAR-RIS. Note that $\mathcal R_{\text{x},1}$ is a monotonically increasing function of  $\mathbf{s}^r_{\text{x}}$ and  it is maximized when $\mathbf{\Theta}^r_{\text{x}}$ is set to co-phase the reflected and  direct paths at $U_1$. Accordingly, the optimal choice of phases that maximizes the objective is given by 
\begin{eqnarray}\label{eqn:optimalreflection}
	\text{Arg}(s^r_{\text{x},m}) = \psi_f - (\psi_{v,m}+\psi_{g_d,m}),\;\; \forall m \in \mathcal M.
\end{eqnarray} 

In (\ref{eqn:ObjectiveP1})-(\ref{eqn:rateconstraints}), the objective and QoS constraint $\mathrm C_5$ are neither convex nor concave functions of the variables. Moreover, the optimization problem becomes a mix-integer non-convex program for the MS protocol due to binary constraint $\mathrm  C_{4}$. Hence, it is difficult to find the optimal solution for the formulated problem. In this context, we propose successive approximation based efficient algorithms to solve (\ref{eqn:ObjectiveP1})-(\ref{eqn:rateconstraints}).

\section{Solution for the Optimization Problem }

In this section, an iterative  strategy is proposed to solve  $\mathrm P1$. First, by using the successive approximation technique, the original problem is  approximated into a more tractable form. Then, efficient algorithms are developed  to design RIS coefficients for the  ES and MS protocols.

\subsection{Reformulation of the Original Problem}
To begin with, we invoke successive approximation technique to tackle the non-concavity of the user rates. In particular, we use Taylor series expansion to  derive lower bounds for $\mathcal R_{\text{x},1}$ and $\mathcal R_{\text{x},2}$ as follows:

\begin{eqnarray}
	\mathcal R_{\text{x},1}
	\geq
	\mathcal R^{lb,i}_{\text{x},1}
	\text{ and }
	\mathcal R_{\text{x},2} 
	\geq	\mathcal R^{lb,i}_{\text{x},2}, 
\end{eqnarray}
where $\mathcal R^{lb,i}_{\text{x},1}$ and $\mathcal R^{lb,i}_{\text{x},2}$ are given in (\ref{eqn:lb_rate1}) and (\ref{eqn:lb_rate2}), respectively, as shown at the top of the next page. 
 The parameters $\psi^i_\text{x}$, ${\kappa^i_\text{x}}$, and $\chi^{i}_{\text{x}}$ in (\ref{eqn:lb_rate1}) and (\ref{eqn:lb_rate2}) are given by $	\psi^i_\text{x} = \sqrt{\bar{\gamma}_1}({f}+(\mathbf s^r_{\text{x},i})^{T}  {\mathbf h}),  
		{\kappa^i_\text{x}} =  {\bar{\gamma}'_2}\left|(\mathbf s^t_{\text{x},i})^{T}\mathbf q\right|^{2}+ 1$,
 and $\chi^{i}_{\text{x}} =  
		\sqrt{\bar{\gamma}_2}(\mathbf s^t_{\text{x},i})^{T}\mathbf z$
where $i$ denotes the iteration index. 
The derivation of (\ref{eqn:lb_rate1}) and (\ref{eqn:lb_rate2}) follows mathematical manipulations similar to those used in \cite[Appendix A]{Chen2019}, and thus omitted here due to limited space.
Note that $\mathcal R^{lb,i}_{\text{x},1}$ and $\mathcal R^{lb,i}_{\text{x},2}$ are concave functions of $\mathbf s^r_\text{x}$ and $\mathbf s^t_\text{x}$. This is particularly important since our reformulated problem iteratively maximizes a lower bound of the original objective of $\mathrm P1$ subject to the constraints. Thus,  we present the  approximated reformulation of $\mathrm P1$ for the ES and MS protocols in the following subsections.

\setcounter{mycounter}{\value{equation}}
\begin{figure*}[!t]
	\addtocounter{equation}{0}
	\begin{small}	
		\begin{eqnarray}		
&&\!\!\!\!\!\!\mathcal R^{lb,i}_{\text{x},1} \!=\! \frac{1}{\ln(2)}\!
\left[
	\ln \left(\! 1 \!+\! \frac{\left|\psi^i_\text{x}\right|^2}
{\kappa^i_\text{x}}\right)
 \!+2\sqrt{\bar{\gamma}_1}\frac{\text{Re}\left[(\psi^i_\text{x} )^H(f + (\mathbf s^r_{\text{x}})^T\mathbf h) \right] }{\kappa^i_\text{x}} \! - \! \frac{\left|\psi^i_\text{x}\right|^2}{\kappa^i_\text{x}(\kappa^i_\text{x}+ \left|\psi^i_\text{x}\right|^2)}
\!\! \left( \bar{\gamma}_1\left|f+(\mathbf s^r_\text{x})^T  \mathbf h \right|^{2}
 + \bar{\gamma}_2\left|(\mathbf s^t_\text{x})^T \mathbf z \right|^2 \right)\! - \!\frac{\left|\psi^i_\text{x}\right|^2}{\kappa^i_\text{x}}
 \right]\label{eqn:lb_rate1}\\
&&\!\!\!\!\!\!\mathcal R^{lb,i}_{\text{x},2} = 	
\frac{1}{\ln(2)}
\left[\ln(1+ |\chi^{i}_{\text{x}}|^2) + 2\sqrt{\bar{\gamma}_2} \text{Re} \left[(\chi^{i}_{\text{x}})^H \left((\mathbf s^t_{\text x})^T \mathbf z -\chi^{i}_{\text{x}}\right)\right]
-\left(\frac{|\chi^{i}_{\text{x}}|^2}{1+|\chi^{i}_{\text{x}}|^2}\right)
\!\left( \bar{\gamma}_2|(\mathbf s^t_{\text{x}})^T \mathbf z|^2
- |\chi^{i}_{\text{x}}|^2 \right)  \right]
\label{eqn:lb_rate2}
		\end{eqnarray}	
	\end{small}	
	\vspace{-2mm}
	\hrulefill
	\vspace{-2mm}
\end{figure*}
\setcounter{equation}{\value{mycounter}}

\subsection{Proposed Solution for the ES Protocol}

Using $\mathcal R^{lb,i}_{\text{E},1}$ and $\mathcal R^{lb,i}_{\text{E},2}$, the $i$th iteration of the reformulated problem can be written as 
\addtocounter{equation}{2}
\begin{subequations}
	\begin{eqnarray}
	\mathrm {P}2:\;
		\underset{\tilde{\mathcal R}_{1},\tilde{\mathcal R}_{2},\mathbf s^k_{\text{E}}\;\; \forall k}{\text{maximize}}&& \!\!\!\! w_1 \tilde{\mathcal R}_{1} + w_2 \tilde{\mathcal R}_{2}\label{eqn:ObjectiveES}\\
	 \;\;\;	\text{subject to} && \nonumber\\
		&&\!\!\!\!\!\!\!\!\!\!\!\!\!\!\!\!\!\!\!\!\!\!\!\!\!\!\!\!\!\!\!\!\!\!\!
		\mathrm C_{1}: 
		\tilde{\mathcal R}_{1} \leq \mathcal R^{lb,i}_{\text{E},1} \text{ and }
		\tilde{\mathcal R}_{2}  \leq \mathcal R^{lb,i}_{\text{E},2} ,
		\label{eqn:rateConstraintES} \\
		&&\!\!\!\!\!\!\!\!\!\!\!\!\!\!\!\!\!\!\!\!\!\!\!\!\!\!\!\!\!\!\!\!\!\!\!
		\mathrm  C_{2}:
		|s^r_{\text{E},m}|^2 + |s^t_{\text{E},m}|^2 \leq 1,\quad \forall m \in \mathcal M,\label{eqn:amplitudeES} \\
		&&\!\!\!\!\!\!\!\!\!\!\!\!\!\!\!\!\!\!\!\!\!\!\!\!\!\!\!\!\!\!\!\!\!\!\!
		\mathrm  C_{3}:
		0\leq  |s^k_{\text{E},m}| \leq 1, \;\;\forall m \in \mathcal M,\label{eqn:amplitudeconstraintES} \\
		&&\!\!\!\!\!\!\!\!\!\!\!\!\!\!\!\!\!\!\!\!\!\!\!\!\!\!\!\!\!\!\!\!\!\!
		\mathrm  C_{4}:
		\mathcal R^{lb,i}_{\text{E},1} \geq \mathcal R^{th}_1 \text{ and }  \mathcal R^{lb,i}_{\text{E},2} \geq \mathcal R^{th}_2,
		\label{eqn:rateconstraintsES}	
	\end{eqnarray}
\end{subequations} 
where $\tilde{\mathcal R}_{1}$ and $\tilde{\mathcal R}_{2}$ are auxiliary variables.
In (\ref{eqn:ObjectiveES})-(\ref{eqn:rateconstraintsES}), the objective and all constraints are convex and hence can be efficiently solved via  CVX; a MATLAB-based software \cite{boyd_vandenberghe_2004}. The complete steps for solving $\mathrm {P}2$ are outlined in Algorithm 1.
\begin{algorithm}\label{algorithm1}
	\caption{\!\!: SCA Algorithm for the ES Protocol}
	\begin{algorithmic}[1]
		\STATE \textbf{Initialize} Set feasible points $\{\mathbf s^k_{\text{E},0}\}$ $\forall k \in\{t,r\}$ (use Remark 1 to set the phase of 	${\mathbf s^r_{\text{E},0}}$). Define the maximum number of iterations $i_{\max}$ and the threshold $\epsilon_1$.

		\STATE \textbf{Iteration $i$}:  
		
		For given $\mathbf s^k_{\text{E},i}$, solve the Problem (\ref{eqn:ObjectiveES})-(\ref{eqn:rateconstraintsES}) and obtain the optimal solution $\mathbf s^{k*}_{\text{E}}$.
		
		\STATE \textbf{Until:} If the fraction increase of the objective value is below  $\epsilon_1 $ or $i_{\max}$ is reached, goto Step 5.
		
		Otherwise, goto Step 4.
		
		\STATE  $\mathbf s^k_{\text{E},i}= \mathbf s^{k*}_{\text{E}}$,  $i=i+1$, and goto Step 2
		
		\RETURN $\mathbf s^{k*}_{\text{E}}$ $ k \in \{t,r\}$.
	\end{algorithmic} 
\end{algorithm}

\noindent \textbf{Remark 2:} Note that the equality constraint in (\ref{eqn:amplitudeconstraintxx}) is relaxed  in (\ref{eqn:amplitudeES}) to make it convex. At the optimal solution, (\ref{eqn:amplitudeES})  satisfies with equality. To demonstrate this, assume that  (\ref{eqn:amplitudeES}) is satisfied with strict inequality. Then, we can always increase $|s^r_{\text{E},m}|$ $\forall m \in \mathcal M$ to satisfy the constraint with equality, which also increases the objective value.

\subsection{Proposed Solution for the MS Protocol}
For the MS protocol, the reformulated problem  can be expressed as
\begin{subequations}
	\begin{eqnarray}
	\mathrm {P}3:\;
		\underset{\tilde{\mathcal R}_{1},\tilde{\mathcal R}_{2},\mathbf s^k_{\text{M}}\;\; \forall k}{\text{maximize}}&& \!\!\!\! w_1 \tilde{\mathcal R}_{1} + w_2 \tilde{\mathcal R}_{2}\label{eqn:ObjectiveMS}\\
		\text{subject to} && \nonumber\\
		&&\!\!\!\!\!\!\!\!\!\!\!\!\!\!\!\!\!\!\!\!\!\!\!\!\!\!\!\!\!\!\!\!\!\!\!
		\mathrm C_{1}: 
		\tilde{\mathcal R}_{1} \leq \mathcal R^{lb,i}_{\text{M},1} \text{ and }
		\tilde{\mathcal R}_{2}  \leq \mathcal R^{lb,i}_{\text{M},2} ,
		\label{eqn:rateConstraintMS} \\
		&&\!\!\!\!\!\!\!\!\!\!\!\!\!\!\!\!\!\!\!\!\!\!\!\!\!\!\!\!\!\!\!\!\!\!\!
		\mathrm  C_{2}:
		|s^r_{\text{M},m}|^2 + |s^t_{\text{M},m}|^2 \leq 1,\quad \forall m \in \mathcal M, \\
		&&\!\!\!\!\!\!\!\!\!\!\!\!\!\!\!\!\!\!\!\!\!\!\!\!\!\!\!\!\!\!\!\!\!\!\!
		\mathrm  C_{3}:
		|s^k_{\text{M},m}| \in \{0,1\},\; \forall m \in \mathcal M ,\label{eqn:amplitudeconstraintMS} \\
		&&\!\!\!\!\!\!\!\!\!\!\!\!\!\!\!\!\!\!\!\!\!\!\!\!\!\!\!\!\!\!\!\!\!\!\!
		\mathrm  C_{4}:
		\mathcal R^{lb,i}_{\text{M},1} \geq \mathcal R^{th}_1 \text{ and }  \mathcal R^{lb,i}_{\text{M},2} \geq \mathcal R^{th}_2,
		\label{eqn:rateconstraintsMS}	
	\end{eqnarray}
\end{subequations} 
where $\mathcal R^{lb,i}_{\text{M},1}$ and $\mathcal R^{lb,i}_{\text{M},2}$ are deduced from (\ref{eqn:lb_rate1}) and (\ref{eqn:lb_rate2}), respectively. 
Due to the binary constraint (\ref{eqn:amplitudeconstraintMS}), the problem $\mathrm P3$ is a mixed-integer non-convex problem which is NP hard in general. To alleviate the complexity, we use a penalty based method in which the non-convex constraint is formulated as a convex term and added to the objective function. To this end, the binary constraint always satisfy $(\delta^k_m)- (\delta^k_m)^2 \geq 0 \;\; \forall k \in \{t,r\}, m \in \mathcal M$, where $\delta^k_{\text{M},m} $ is the amplitude of $s^k_{\text{M},m}$. Here, the equality constraint holds if and only if $\delta^k_{\text{M},m} $ is a binary variable, i.e., $0$ or $1$. By using the Taylor series expansion, an upper bound can be derived as  
\begin{eqnarray}\label{eqn:beta}
	 \delta^{k}_{\text{M},m} \!-\! \left(\delta^{k}_{\text{M},m}\right)^2
	 \!&\leq&\! \!\! \left(\delta^{k,i}_{\text{M},m}\right)^2 \!+ \left(1- 2 \delta^{k,i}_{\text{M},m}\right) \delta^{k}_{\text{M},m} \nonumber\\
	 &=&\!\!\! \Pi\left(\delta^{k}_{\text{M},m}, \delta^{k,i}_{\text{M},m}\right),\;  \forall m \in \mathcal M,
\end{eqnarray}
where $\forall k \in \{t,r\}$ and $\delta^{k,i}_{\text{M},m}$ is a given point in the $i$th iteration of SCA. By adding the upper bound for the binary constraint as a penalty term, $\mathrm {P}3$ can be reformulated as 
\begin{subequations}
	\begin{eqnarray}
\mathrm P4:		\underset{\tilde{\mathcal R}_{1},\tilde{\mathcal R}_{2}, \delta^{k}_{\text{M},m},  s^k_{\text{M}}\;\; \forall k}{\text{maximize}}&& \!\!\!\! w_1 \tilde{\mathcal R}_{1} + w_2 \tilde{\mathcal R}_{2} 
-\mu \hat{\Pi}
		\label{eqn:ObjectiveMS2}\\
		\text{subject to} && \nonumber\\
		&&\!\!\!\!\!\!\!\!\!\!\!\!\!\!\!\!\!\!\!\!\!\!\!\!\!\!\!\!\!\!\!\!\!\!\!
		(\ref{eqn:rateConstraintMS}), (\ref{eqn:rateconstraintsMS})\\
		&&\!\!\!\!\!\!\!\!\!\!\!\!\!\!\!\!\!\!\!\!\!\!\!\!\!\!\!\!\!\!\!\!\!\!\!
		\mathrm  C_{2}:
		(\delta^r_{\text{M},m})^2 + (\delta^r_{\text{M},m})^2 \leq 1,\;\; \forall m \in \mathcal M,\quad \\
		&&\!\!\!\!\!\!\!\!\!\!\!\!\!\!\!\!\!\!\!\!\!\!\!\!\!\!\!\!\!\!\!\!\!\!\!
		\mathrm  C_{3}:
		\delta^k_{\text{M},m}\geq|s^k_{\text{M},m}|,\; \forall m \in \mathcal M ,\label{eqn:amplitudeconstraintMS2} 
	\end{eqnarray}
\end{subequations} 
where $\hat{\Pi} = \sum_{m=1}^M\sum_{k\in \{t,r\}} \Pi\left(\delta^{k}_{\text{M},m}, \delta^{k,i}_{\text{M},m}\right)$ and $\mu$ is the penalty factor. The objective and constraints of (\ref{eqn:ObjectiveMS2})-(\ref{eqn:amplitudeconstraintMS2}) are convex and can be solved by applying the SCA technique using CVX. Moreover, our extensive simulations reveal that (\ref{eqn:amplitudeconstraintMS2}) holds equality at the optimal solution. Details of the proposed SCA algorithm are presented in Algorithm 2, which consists of two loops. The outer-layer iteration is used to update the penalty factor  while the inner-layer iteration is used to solve $\mathrm P4$ via SCA. For the outer-layer, the termination condition is 
\begin{eqnarray}
	\max\{|s^{k}_{\text{M},m}|-|s^{k}_{\text{M},m}|^2, \forall k \in \{t,r\}, m \in \mathcal M\}\leq \epsilon_2,
\end{eqnarray}
where $\epsilon_2>0$ is the predefined accuracy with which constraint (\ref{eqn:amplitudeconstraintMS}) is met.
\begin{algorithm}\label{algorithm2}
	\caption{\!\!: Penalty-Based Iterative Algorithm for the MS Protocol}
	\begin{algorithmic}[1]
		\STATE \textbf{Initialize} Set feasible points $\{\mathbf s^k_{\text{M},0} \}$ $\forall k \in\{t,r\}$ (set the phase of 	${\mathbf s^r_{\text{M},0}}$ as per Remark 1). Define the maximum number of iterations $i_{\max}$, the thresholds $\epsilon_1$, $\epsilon_2$, and the penalty parameters $\mu,\omega$.

		\STATE \textbf{Outer-layer}: 
		
		Set the iteration index $i=0$ for inner-layer.   
		
		\STATE \hspace{3mm}\textbf{Inner-layer}: Iteration $i$
		
		\hspace{3mm} For given $\{\mathbf s^k_{\text{M},i}\}$, solve the problem (\ref{eqn:ObjectiveMS2})- (\ref{eqn:amplitudeconstraintMS2}) and
		
		\hspace{3mm} obtain the optimal solution $\mathbf s^{k*}_{\text{M}}$.
		
		\STATE \hspace{3mm}\textbf{Until} If the fraction increase of the objective value is 
		
		\hspace{3mm} below  $\epsilon_1 $ or $i_{\max}$ is reached, goto Step 6.
		
		\hspace{3mm} Otherwise, goto Step 5.
		
		\STATE \hspace{3mm}   $\mathbf s^k_{\text{M},i}= \mathbf s^{k*}_{\text{M}}$,  $i=i+1$, and goto Step 3
		
		\STATE  Update $\{\mathbf s^k_{\text{M},0}\}$ with $\mathbf s^{k*}_{\text{M}}$ and $\mu = \omega \mu$, goto Step 2
		
		\STATE \textbf{Until} the constraint violation falls below $\epsilon_2 >0$.
		
		\RETURN $\mathbf s^{k*}_{\text{M}}$  $\forall k \{t,r\}$.
	\end{algorithmic} 
\end{algorithm}

\subsection{Description of Algorithms}

\subsubsection{Convergence Analysis}
The Algorithm 1 and Algorithm 2 repeatedly solve $\mathrm P2$ and $\mathrm P4$ until the objective of the original problem $\mathrm P1$ converges. In particular, denoting the objective values of $\mathrm P1$  and $\mathrm P2 $ as $G(\mathbf s^r_{\text{E}},\mathbf s^t_{\text{E}})$ and $F(\mathbf s^r_{\text{E}},\mathbf s^t_{\text{E}})$, it follows that $G(\mathbf s^r_{\text{E},i+1},\mathbf s^t_{\text{E},i+1})\geq F(\mathbf s^r_{\text{E},i+1},\mathbf s^t_{\text{E},i+1})\geq G(\mathbf s^r_{\text{E},i},\mathbf s^r_{\text{E},i})$. The above inequality holds because the objective of $\mathrm P2$ lower bounds that of $\mathrm P1$. A similar argument can be made in the case of Algorithm 2. Hence, convergence is guaranteed for both the algorithms.

\subsubsection{Feasibility and Initialization} 
Due to the stringent QoS requirements imposed by large $\mathcal R^{th}_1$ and $\mathcal R^{th}_2$ values, $\mathrm {P}2$ and $\mathrm {P}4$ can be infeasible. 
Hence, prior to solving the problem ($\mathrm {P}2$ or $\mathrm {P}4$), a feasibility-check is performed. The goal of the feasibility-check is to find a solution that satisfies all constraints of the problem. If no solution exists, both user rates are set to zero, i.e., $\mathcal R_{\text{x},1} = 0, \mathcal R_{\text{x},2} = 0$. Otherwise, the feasibility-check solution is assigned as the initial value for  $\{\mathbf s^k_{\text{x}} \}$, i.e., $\{\mathbf s^k_{\text{x},0} \}$ $\forall k \in\{r,t\}$.

\subsubsection{Complexity Analysis}
The proposed algorithms iteratively solve a series of convex optimization problems. In particular, both $\mathrm P2$ in (\ref{eqn:ObjectiveES})-(\ref{eqn:rateconstraintsES}) and  $\mathrm P3$ in (\ref{eqn:ObjectiveMS2})-(\ref{eqn:amplitudeconstraintMS2}) are quadratic-constrained quadratic programming (QCQP) problems, which are  forms of convex optimization problems \cite{boyd_vandenberghe_2004}. These convex optimization problems can be efficiently solved via the CVX software.
CVX utilizes the SDP3 solver, which adopts interior-point methods to solve convex problems. Hence the complexity of both $\mathrm P2$ and $\mathrm P3$ is $\mathcal O[(2M)^{3.5}]$. Thereby, the overall complexity of Algorithm 1 is  $\mathcal O[I_{E}(2M)^{3.5}]$, where $I_E$ is the number of iterations required for convergence. Similarly, the complexity of Algorithm 2 is $\mathcal O[I_oI_{i}(2M)^{3.5}]$, where $I_i$ and $I_o$ denote the number of inner and outer iterations required for convergence, respectively.

\section{Simulation Results}
In this section, simulation results are presented to validate the performance of the proposed STAR-RIS assisted FD communication system. In our simulated setup,  the large-scale fading coefficients are modeled as $-\zeta_{Q}[\text{dB}]= \zeta_0 + 10 \nu \log_{10}(d)$ for $Q\in\{g,u,v,f\}$, where $\zeta_0 = 42$\,dB is a reference path loss, $\nu = 3.5$ is the path loss exponent, and $d$ is the communication link distance (in meters). Here, the distances from AP-to-RIS, AP-to-$U_1$, RIS-to-$U_1$ and $U_2$ are $80$\,m, $100$\,m $30$\,m, and $20$\,m, respectively. Moreover, the channel parameters are set to $m_f =1, m_g =4, m_u = 3$, and $m_u = 2$. Unless otherwise specified, other simulation parameters are set as follows: channel bandwidth $10$\,MHz, noise power density $-174$\,dBm/Hz, noise power $\sigma^2_1 =\sigma^2_1 = -84$\,dBm, tolerance for objectives $\epsilon_1 = \epsilon_2 = 0.001$, and weight factors $w_1 =0.7$ and $w_2 = 0.3$. The average transmit signal-to-noise ratio (SNR) for $U_1$ and $U_2$ are assigned as $\bar{\gamma}_1 = \bar{\gamma}$ and $\bar{\gamma}_2 = \bar{\gamma}/2$. For simplicity, the QoS requirements are set as $R^{th}_i = \log_2(1+\kappa \bar{\gamma}_i)$ for $\kappa =0.1$ for $i=\{1,2\}$. The reflection and transmission coefficients at the STAR-RIS are initialized by uniformly and randomly selecting the amplitude and phase shift of each coefficient \cite{Peng2021}.  

\subsection{Baseline Schemes}
To highlight the benefits of STAR-RIS assisted FD communication, we consider two baselines, namely, the conventional RIS scheme and HD scheme. In the conventional RIS scheme, the AP serves $U_1$ and $U_2$ via a composite intelligent surface consists of reflecting-only and transmitting-only elements, each with $M_r$ and $M_t$ elements, such that $M_t +M_r = M$. This scheme can be considered as a special case of the MS protocol in which the first $M_r$ elements operate in reflection mode while the next $M_t$ elements operate in transmission mode. In the HD scheme, we consider an HD AP, which serves $U_1$ and $U_2$ in two orthogonal time slots $\lambda_1$ and $\lambda_2$, respectively, such that $\lambda_1+\lambda_2 =1$. In this scheme, both SI and inter-user interference  vanish. Moreover, the transmit powers should be scaled by $1/\lambda$ to properly compare with the FD system. Therefore, the rate expressions of the HD system can be written as 
\begin{eqnarray}
	\mathcal R_{\text{H},1} \! &=& \! \!
	\lambda_1\log_2 \left(\! 1 + {\bar{\gamma}_1\left|f+(\mathbf s^r_\text{H})^T  \mathbf h \right|^{2}}
	\big/{\lambda_1}\right),\label{eqn:user_1_rateHD}  \text{ and }\\
	\mathcal R_{\text{H},2} &=& \lambda_2\log_2  \left(1+\bar{\gamma}_2\left|(\mathbf s^t_\text{H})^T \mathbf z \right|^2\big/\lambda_2 \right),\label{eqn:user_2_rateHD}
\end{eqnarray}
respectively, where $\mathbf s^r_\text{H}$ and $\mathbf s^t_\text{H}$ are the reflection and transmission coefficients at the RIS for the HD system and can be derived in closed-form as
\begin{eqnarray}\label{eqn:optimalreflectionHD}
	&&\!\!\!\!\!\!\!\!\!\!\!
	|s^r_{\text{H},m}| = 1,\text{ and }  \text{Arg}(s^r_{\text{H},m}) = \psi_f - (\psi_{v,m}+\psi_{g_d,m}),\\
	&&\!\!\!\!\!\!\!\!\!\!\!
	|s^t_{\text{H},m}| = 1,\text{ and }  \text{Arg}(s^t_{\text{H},m}) = -(\psi_{u,m}+\psi_{g_u,m}),
\end{eqnarray}
respectively, for $\forall m \in \mathcal M$. To facilitate a fair comparison between the proposed FD system and the HD counterpart, we optimize $\lambda_1$ and $\lambda_2$ time-slots as
\begin{subequations}
	\begin{eqnarray}
		\mathrm {P}5:\;
		\underset{\lambda_1,\lambda_2}{\text{maximize}}&& \!\!\!\! w_1 {\mathcal R}_{\text{H},2} + w_2 {\mathcal R}_{\text{H},2}\label{eqn:ObjectiveHD}\\
		\text{subject to} && \nonumber\\
		&&\!\!\!\!\!\!\!\!\!\!\!\!\!\!\!\!\!\!\!\!
		\mathrm C_{1}: \lambda_1+\lambda_2=1,
		\label{eqn:rateConstraintHD} \\
		&&\!\!\!\!\!\!\!\!\!\!\!\!\!\!\!\!\!\!\!\!
		\mathrm  C_{2}:
		0\leq \lambda_1 \leq 1,\;	0\leq \lambda_2 \leq 1,\label{eqn:amplitudeconstraintHD} \\		&&\!\!\!\!\!\!\!\!\!\!\!\!\!\!\!\!\!\!\!\!
		\mathrm  C_{3}:
		\mathcal R_{\text{H},1} \geq \mathcal R^{th}_1 \text{ and }  \mathcal R_{\text{H},2} \geq \mathcal R^{th}_2.
		\label{eqn:rateconstraintsHD}
	\end{eqnarray}
\end{subequations} 
 $\mathrm {P}5$ is a convex problem and can be easily solved using convex optimization tools such as CVX \cite{boyd_vandenberghe_2004}.

\begin{figure}[!t]\centering\vspace{-0mm}
	\includegraphics[width=0.4\textwidth]{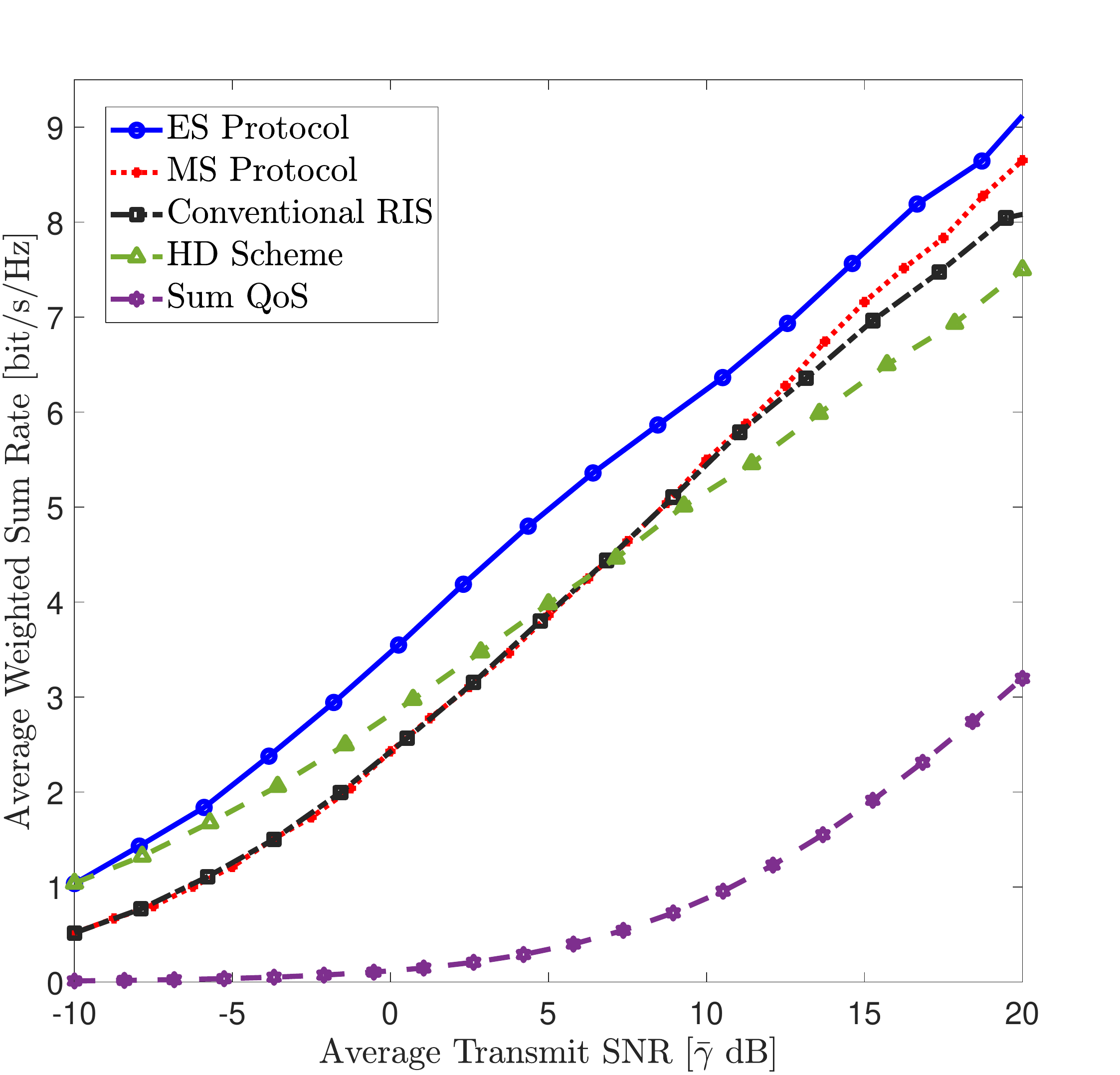}
	\vspace{-3mm}
	\caption{The average weighted sum rate versus average transmit SNR for $M=64$. }
	\label{fig:Figure1V1}\vspace{-5mm}
\end{figure}

\begin{figure}[!t]\centering\vspace{-0mm}
	\includegraphics[width=0.4\textwidth]{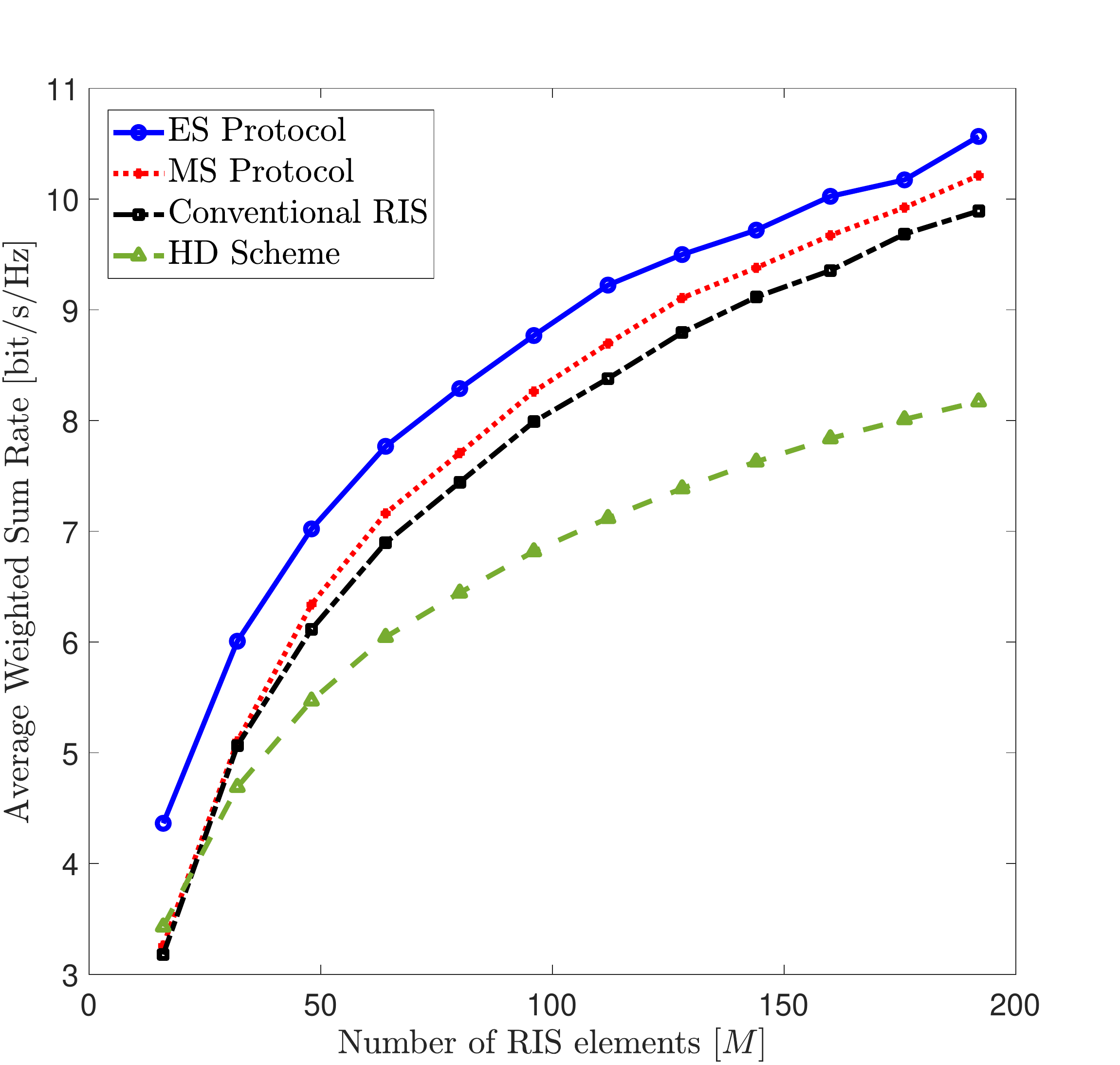}
	\vspace{-3mm}
	\caption{The average weighted sum rate versus  $M$ for $\bar{\gamma} = 15$\,dB. }
	\label{fig:Figure2}\vspace{-5mm}
\end{figure}

In Fig. \ref{fig:Figure1V1}, weighted sum rates of the ES and MS protocols are plotted as functions of $\bar{\gamma}$ for $M=64$ by using the  Algorithm 1 and Algorithm 2, respectively. The weighted QoS sum rate requirement, ($w_1 \mathcal R^{th}_1+w_2 \mathcal R^{th}_2$) is also plotted for clarity. Fig. \ref{fig:Figure1V1} shows that the ES protocol outperforms the MS protocol throughout the considered SNR regime while satisfying the QoS requirement. The average weighted sum rate for baseline schemes (i.e., conventional RIS and HD scheme) are also plotted for comparison purposes. The STAR-RIS ES protocol has a superior sum rate performance than the conventional RIS, while the STAR-RIS MS protocol provides a similar performance for $M = 64$ (also see Fig \ref{fig:Figure2}).  
The curve pertaining to the HD scheme is plotted by solving $\mathrm P5$. Since HD scheme is not affected by the SI and inter-user interference, it outperforms the MS protocol at low SNR. Nevertheless, for moderate to  high SNR values, the sum rates of the ES and MS protocols surpass the baseline schemes. Thus, Fig. \ref{fig:Figure1V1} reveals that the integration of STAR-RISs to the existing FD communication systems will boost the sum rate while meeting stringent user rate requirements.

 In Fig. \ref{fig:Figure2}, the average weighted sum rate performance of the STAR-RIS protocols and baseline schemes is plotted for different number of RIS elements. From Fig. \ref{fig:Figure2}, it is observed that the rate performance of both ES and MS protocols increases with diminishing returns. For instance, when the number elements are  doubled from $M=32$ to $64$ and from $M=64$ to $128$, the increments of weighted sum rate for ES (MS) protocol   
  are 29.3\% (40.3\%) and 22.3\% (27.1\%), respectively. Moreover, STAR-RIS protocols outperform both baseline schemes. In particular, the MS protocol and conventional RIS scheme provide a similar performance for fewer elements, i.e., $M\leq48$. The former slightly outperforms the latter when the number of elements increases, i.e., $M>48$. This performance gain is due to the flexibility introduced by the MS protocol to intelligently determine the operating mode of each RIS element instead of pre-determined operating modes. Moreover, the ES protocol outperforms the MS protocol and both baseline schemes. This observation indicates the importance of intelligently selecting the amplitudes of reflection and transmission coefficients to boost the performance of STAR-RIS assisted FD communication systems.

\vspace{-2mm}
\section{Conclusion}
 The rate performance of a STAR-RIS assisted FD communication system was investigated. In particular, the reflection and transmission coefficients of the STAR-RIS have been optimized to maximize the weighted sum of uplink and downlink user rates subject to QoS requirements and unit modulus constraints of the STAR-RIS elements. Two  SCA-based algorithms were proposed to efficiently solve the original non-convex optimization problems for the ES and MS protocols. The performance of the proposed system designs was compared with the conventional RIS and the HD counterparts via simulations, where it was observed that: (i) ES protocol has better performance than the MS protocol, (ii) both the ES and MS protocols outperform the baseline schemes as the number of STAR-RIS elements are increased.

\bibliographystyle{IEEEtran}
\bibliography{IEEEabrv,Reference2}

\end{document}